\documentclass[FBSedit,FBSmath,ecsub]{FBSsuppl}
\usepackage{amsfonts}
\usepackage{amssymb}
\usepackage{epsfig}

\title{The Nucleon-Nucleon Problem in Quark Models}
\author{Fl. Stancu\thanks{
\textit{E-mail address:} fstancu@email.ulg.ac.be}}
\institute{Institute of Physics B.5, University of Li\`ege,
Sart Tilman, B-4000 Li\`ege 1, Belgium}
\begin{document}

\maketitle

\begin{abstract}
%The NN problem has a long history, about 70 years old.
%However 
%the emphasis will be here only on calculations to test aspects of QCD
%via quark models.
In the first part we summarize the status of the nucleon-nucleon (NN) problem 
in the context of Hamiltonian 
based constituent quark models and present results for the 
$\ell = 0$ phase shifts obtained from the Goldstone-boson exchange 
model by applying the resonating group method.
The second part deals with the construction of local shallow and
deep equivalent potentials based on a Superymmetric
Quantum Mechanics approach.
\end{abstract}

\section{Introduction}

%%%%%%%%%%%%%%%%%%%%%%%%%%%%%%%%%%%%%%%%%%%%%%%%%%%%%%%%%%%%%%%%%%%%%%%%
%
%
%At the previous Few-Body Conference recent progress in the construction
%of phenomenological nucleon-nucleon (NN) potentials has been reviewed 
%\cite{EVORA}.
% The current
%status of high-precision, charge dependent nucleon-nucleon potentials
%\cite{STOKS93,ARGONNE,BONN_NEW}
%has been presented. These potentials are constructed in the spirit of
%meson theory and fit the data typically with a $\chi^2$/datum $\approx$ 1,
%while the best models of the 80's \cite{PARIS,BONN} fit the data with
%$\chi^2$/datum $\approx$ 2. 
%In these studies, either the concept of boson exchange is illegitimately
%extended to very short internucleon distances \cite{STOKS93,BONN} or
%a purely phenomenological model is explicitly adopted for these distances
%\cite{PARIS}. This remark together with the fact that the high-precision
%potentials have about 45 parameters to fit, raises the question to
%what extent can one understand the short range behaviour of the NN
%interaction potential from a microscopical point of view, where the
%quark structure of nucleons is taken into account.
%%%%%%%%%%%%%%%%%%%%%%%%%%%%%%%%%%%%%%%%%%%%%%%%%%%%%%%%%%%%%%%%%%%%%%%
The strong interaction between two nucleons is the basic ingredient
of nuclear physics. Since about 25 years there have been many efforts 
to derive the nuclear forces from the underlying theory of strong
interactions, the Quantum Chromodynamics (QCD). As QCD cannot be
solved exactly in the low energy regime,  QCD-inspired
models have been used both in baryon spectroscopy and the
NN problem.
We first briefly describe the situation  
in the framework of
Hamiltonian based quark models. Next, we present 
recent progress obtained from using the Goldstone-boson exchange model.

After some pioneering work in the late '70
\cite{LIBERMAN}
a major breakthrough in the microscopical derivation of the NN interaction
has been achieved at the begining of the 80's
with the work of Oka and Yazaki \cite{OY} who used the
resonating group method to derive phase shifts
from the one-gluon exchange model, the work of Harvey \cite{HARVEY} 
on the symmetries of the most important six-quark states and
the work of Golli, Rosina and collaborators \cite{CGMR}
on local effective nucleon-nucleon interactions.

   Presently there are several review papers available on the subject 
\cite{OY2}. They show that 
the following steps are important:
1) the choice of the quark model,
2) the choice of the six-quark basis states,
3) the method to calculate the phase shifts.
A challenge is to describe both the nucleon, as a three-quark system, and
the NN interaction, as a six-quark problem, using the same quark model.

\section{Quark models}

The quark models used so far in the NN problem are mostly nonrelativistic.
Studies based on the one-gluon exchange (OGE) model explained the
short-range repulsion as due to the color-magnetic 
interaction combined with quark interchange between $q^3$ clusters.
But in order to reproduce the scattering data some extra medium-
and long-range repulsion was necessary. This was added
at the baryon level. That is why more consistent models, called
hybrid, have subsequently been introduced \cite{HYBRID}
where all interactions are introduced at the quark level. In
such models the short-range repulsion is still attributed to the OGE 
interaction and the medium- and long-range attraction is due to
scalar and pseudoscalar meson exchange.
To our knowledge, these models have problems in fitting 
the nucleon resonances and the NN interaction with the same parameters.

      Here we present results derived from the Goldstone-boson exchange
(GBE) model which has been succesful in describing the baryon properties
\cite{PLESSAS}. The pseudoscalar meson exchange interaction between quarks,
which is 
spin and flavour dependent, contains both a long and a short range part,
appropriate for the NN problem. In this work we use the parametrization
of ref. \cite{GRAZ}. We employ the resonating group method 
to calculate the $\ell=0$ phase shifts
by using the technique proposed
by Kamimura \cite{KAMIMURA}.

%%%%%%%%%%%%%%%%%%%%%%%%%%%%%%%%%%%%%%%%%%%%%%%%%%%%%%%%%%%%%%
\section{Phase shifts in the GBE model}

\begin{figure}[hbt]
\begin{center}
\epsfig{file=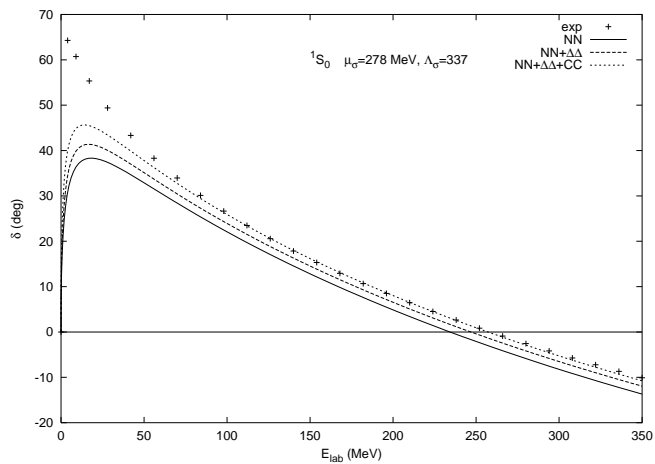,width=250pt}
\caption{\label{Fig. 1} The $^1S_0$ NN scattering phase shift.
The coupled channel calculations are from Ref. \cite{BARTZ}.
The crosses are the Nijmegen fit \cite{STOKS93} to the data.}
\end{center}
\end{figure}

%%%%%%%%%%%%%%%%%%%%%%%%%%%%%%%%%%%%%%%%%%%%%%%%%%%%%%%%%%%%%%%%%%%%%%
%\subsection{The role of the tensor force}

The phase shifts obtained in Ref. \cite{BS1} from the GBE model also imply 
that the NN potential is strongly repulsive at short range. This means
that the repulsion can equally be explained as due to a flavour-spin
interaction combined with quark interchange.
The repulsion obtained from the GBE model is somewhat stronger
than that produced by the OGE model .
However, to reproduce the experimental $^1S_0$ phase shift
a certain amount of middle-range attraction was necessary. This
has been provided by the addition of a scalar meson exchange interaction
of the form \cite{BS2}
\begin{equation}\label{SCALAR}
V_{\sigma}=-\frac{g_{\sigma q}^2}{4\pi}
(\frac{e^{-\mu_{\sigma}r}}{r}-\frac{e^{-\Lambda_{\sigma}r}}{r})\ 
\end{equation}
where we chose  $\frac{g_{\sigma q}^2}{4\pi} = 
\frac{g_{\pi q}^2}{4\pi} = 1.24$
, $\mu_{\sigma} = 278$ MeV, $\Lambda_{\sigma} = 337$ MeV.
Interestingly, the fit to the data favours  $\mu_{\sigma} = 2 m_{\pi}$,
consistent with the findings of Ref. \cite{GR}.
Fig. 1 shows that the addition of (\ref{SCALAR}) leads to a good agreement 
with the data over a large energy interval,
the best result being obtained with three coupled
channels NN+$\Delta\Delta$+CC. The quality of the baryon spectrum remains
unchanged after the addition of (\ref{SCALAR}). 
%%%%%%%%%%%%%%%%%%%%%%%%%%%%%%%%%%%%%%%%%%%%%%%%%%%%%%%%%%%%%%%%%%
\begin{figure}[hbt]
\begin{center}
\epsfig{file=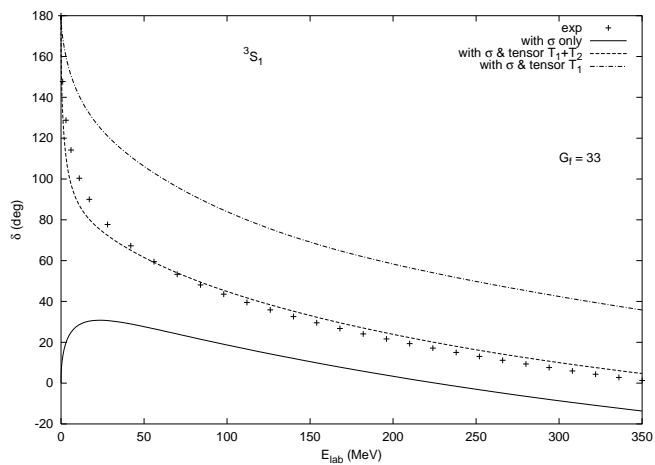,width=250pt}
\caption{\label{Fig. 2} The $^3S_1$ NN scattering phase shift:  
with $\sigma$-meson exchange only (full line),
%where $\mu_{\sigma}$= 278 MeV and $\lambda_{\sigma}$ = 337 MeV,
with $\sigma$-meson exchange + full tensor
interaction with (\ref{RGMtensorpotential}) with $G_f$ = 33 (dotted line),
with $\sigma$-meson exchange
+ 
%part $T_1$ (\ref{RGMtensorpotentialT1}) 
the first term of the tensor interaction (\ref{RGMtensorpotential})  
only, still with $G_f$ = 33 (dot-dashed line).
The crosses are the
Nijmegen fit \cite{STOKS93} to the data.}
\end{center}
\end{figure}

%The phase shifts \cite{BS1} derived from the GBE model indicate that the
%NN potential is strongly repulsive at short range. This means
%that the repulsion ca equally be explained as due to a flavour-spin
%interaction combined with quark interchange.
%The repulsion obtained from the GBE model is somewhat stronger
%than that produced by the OGE model.

%However, to reproduce the experimental $^1S_0$ phase shift
%a certain amount of middle range attraction was necessary. This
%has been provided by the addition of a scalar meson exchange interaction,
%consistent with the pseudoscalar exchange \cite{BS2}
%\begin{equation}
%V_{\sigma}=-\frac{g_{\sigma q}^2}{4\pi}
%(\frac{e^{-\mu_{\sigma}r}}{r}-\frac{e^{-\Lambda_{\sigma}r}}{r})\ 
%\end{equation}
%with $\frac{g_{\sigma q}^2}{4\pi} = 
%\frac{g_{\pi q}^2}{4\pi} = 1.24$
%, ~~~ $\mu_{\sigma}=278$ MeV, ~~~$\Lambda_{\sigma} = 337$ MeV.
%As seen from Fig. 1 a good agreement with the data
%is obtained over a large energy interval. It turned out that the
%best fit was obtained by taking the scalar meson  mass $m_{\sigma}$
%equal to two times the pion mass, as on the chiral circle. 
%It is important to note that the scalar meson exchange does not alter 
%the good quality of the baryon spectrum.

Besides a spin-spin, the pseudoscalar meson exchange gives rise
to a tensor term as well
\begin{eqnarray}\label{RGMtensorpotential}
V^T_{\gamma}(r_{ij})&=&G_f\ \frac{g^2_{\gamma}}{4 \pi}\frac{1}{12 m_i m_j}
\times \nonumber \\
& &\left\{ \mu^2_{\gamma}(1+\frac{3}{\mu_{\gamma}r}+
\frac{3}{\mu^2_{\gamma}r^2})\frac{e^{-\mu_{\gamma}r}}{r}
-\Lambda^2_{\gamma}(1+\frac{3}{\Lambda_{\gamma}r}+
\frac{3}{\Lambda^2_{\gamma}r^2})\frac{e^{-\Lambda_{\gamma}r}}{r}\right\}
\nonumber\\
%&&\nonumber\\
\end{eqnarray}
where $\gamma = \pi,\eta$ and $\eta '$.
%To describe the $^3S_1$ phase shift it was
%necessary to add this contribution to the Hamiltonian \cite{GRAZ}. 
It was found \cite{BARTZ} that the tensor term used in baryon
spectroscopy 
%\cite{TENSOR} 
with $G_f$ = 1 had to have 
$G_f$ = 33 in order to fit the experimental 
$^3S_1$ phase shift.  This is shown in Fig. 2.
The tensor term (\ref{RGMtensorpotential}) has a usual Yukawa type part,
depending on the pseudoscalar meson masses $\mu_{\gamma}$ and a regularized
part, containing cut-off parameters $\Lambda_{\gamma}$.
%\begin{eqnarray}\label{RGMtensorpotential}
%V^T_{\gamma}(r_{ij})&=&G_f\ \frac{g^2_{\gamma}}{4 \pi}\frac{1}{12 m_i m_j}
%\times \nonumber \\
%& &\left\{ \mu^2_{\gamma}(1+\frac{3}{\mu_{\gamma}r}+\frac{3}{\mu^2_{\gamma}r^2})\frac{e^{-\mu_{\gamma}r}}{r}-\Lambda^2_{\gamma}(1+\frac{3}{\Lambda_{\gamma}r}+\frac{3}{\Lambda^2_{\gamma}r^2})\frac{e^{-\Lambda_{\gamma}r}}{r}\right\}\nonumber\\&&\nonumber\\
%\end{eqnarray}
%%\noindent 
%$\gamma = \pi,\eta$ and $\eta '$
If the latter term is removed, 
the attraction increases. Then a smaller factor, $G_f$=12,
is required to reproduce the data.
Further details of these studies can be found in Ref. \cite{BARTZ}.
It would certainly be useful to search for alternative parametrizations
of the GBE model which could better fit the NN phase shifts.
 %%%%%%%%%%%%%%%%%%%%%%%%%%%%%%%%%%%%%%%%%%%%%%%%%%%%%%%%
\section{SUSY approach to local phase shift equivalent potentials.}
\label{secstyle}

A microscopic derivation of the NN interaction, as above, leads to a
nonlocal potential. This is a consequence of the complex structure of
the interacting nucleons. As in the $\alpha-\alpha$ scattering \cite{SAITO},
the wave function of this nonlocal potential
presents a node in the $l=0$ partial wave \cite{BS1}.
It means that if one would try to describe 
the interaction between two nucleons by an equivalent local potential,
this potential would have an extra $l=0$ state. This is an 
unphysical
state known as \textit{Pauli forbidden state}. The mathematical 
relation
between such \textit{deep} potentials and 
phenomenological \textit{shallow} potentials (no
bound state), as e. g. the Reid soft core
potential \cite{REID}, can interestingly and
succesfully be described through a Supersymmtric (SUSY)
Quantum Mechanics approach. The procedure has been originated by Sukumar
\cite{SUKUMAR} and applied 
as a two-step SUSY transformation to the NN scattering by Sparenberg and Baye,
see e.g. ref. \cite{BAYE}.

\begin{figure}[hbt]
\begin{center}
\epsfig{file=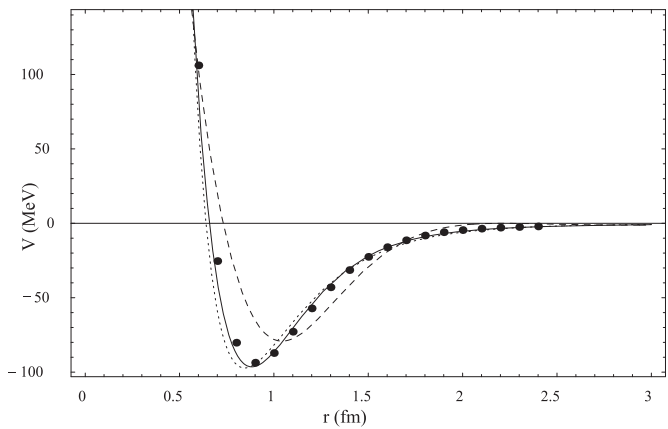,width=250pt}
\caption[]
%{This is an example for an {\protect\textsc{PostScript}} plot. 
%The macro generates the corresponding {\protect\texttt{special}} calls 
%for the printer driver}
{The shallow potential $V_6$ (solid line) compared to
Reid68 \cite{REID} (dotted line),  Baye \& Sparenberg \cite{BAYE}
(dashed line) and Reid93 \cite{STOKS93} (dots) potentials.} 
\end{center}
\end{figure}

\begin{figure}[hbt]
\begin{center}
\epsfig{file=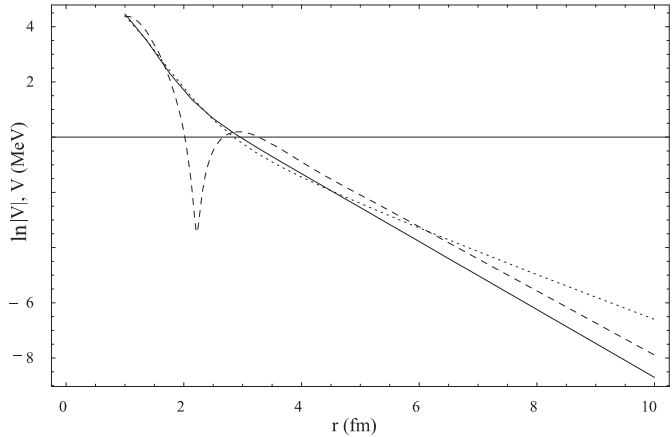,width=250pt}
\caption{Asymptotic behaviour
of the absolute value of  $V_6$ (solid line), 
Reid68  
\protect\cite{REID}
(dotted line),
Baye \& Sparenberg
\protect\cite{BAYE}
(dashed line) in natural logarithmic scale.}
\end{center}
\end{figure}

Here we present an alternatively new procedure \cite{SS} based on 
phase equivalent chains of Darboux (or SUSY) transformations 
at fixed $\ell$. These chains
contain N succesive transformations instead of two, as in ref. \cite{BAYE}.
For potentials related by a chain of transformations we derive an analytic 
expression for the Jost function and hence for the phase shift. They
both contain N parameters, related to the N
unphysical eigenvalues of a starting Hamiltonian, used
to construct the chain. It is convenient to 
choose this Hamiltonian as the free particle Hamiltonian.  
These parameters are also poles of the S-matrix and can be fixed by a fit
to the experimental phase shift.
They can be distributed between poles and zeros of a Jost
function and each distribution corresponds to a different potential. 
We applied this method to derive a shallow and a family of 
deep  phase equivalent potentials for the $l=0$ neutron-proton scattering.
The $^1S_0$ phase shift has optimally been fitted with N = 6 S-matrix poles.
The shallow potential, denoted by $V_6$, can be seen in Fig. 3.
 It is very close to the
Reid soft core potential, denoted by Reid68, and also close to its updated 
version, called Reid93 \cite{STOKS93}. The major improvement over the results
of Ref. \cite{BAYE} can be seen in Fig. 2. The unwanted oscillations
in the potential tail  \cite{BAYE} have disappeared so that the behaviour
of  $V_6$ is consitent
with Yukawa's theory. A deep phase equivalent potential, accomodating a
(Pauli forbidden) bound state, was found by addind two more poles. 
This is the supersymmetric partner of $V_6$.
Varying the available parameters we brought this potential
close to that of Ref. \cite{KUKULIN},
inspired from microscopic calculations
as the sum of a Gaussian plus a Yukawa potential tail.

Thus the use of chains of Darboux (SUSY) transformations at
fixed $\ell$ provides a poweful method
for getting shallow and deep phase equivalent potentials for 
$ \ell = 0 $ partial waves.
Studies of $l \neq 0$ phase equivalent potentials 
based on $\ell$-changing Darboux transformations are underway.

%%%%%%%%%%%%%%%%%%%%%%%%%%%%%%%%%%%%%%%%%%%%%%%%%%%%%%%%%%%%%%%%%%%%%%

\begin{acknowledge}
%I am grateful to my collaborators, Daniel Bartz, who worked hard for 
%his PhD thesis and to Boris Samsonov who introduced me to Supersymmetric
%Quantum Mechanics.
I am grateful to my collaborators, Daniel Bartz and Boris Samsonov
without whom this research could not have been accomplished.
\end{acknowledge}

%%%%%%%%%%%%%%%%%%%%%%%%%%%%%%%%%%%%%%%%%%%%%%%%%%%%%%%%%%%%%%%%%%%%%%

\end{document}